\documentclass{svproc}
\usepackage{graphicx}
\usepackage{bm}
\usepackage{epsfig}
\usepackage{graphics}
\usepackage{amsmath}
\usepackage{amssymb}
\usepackage{latexsym}
\usepackage{slashed,cite}
\usepackage{upgreek}
\usepackage[l]{floatflt}

\usepackage{url}

\def\eec{\end{center}}
\def\bec{\begin{center}}
\newcommand{\eem}{\end{matrix}}
\newcommand{\bem}{\begin{matrix}}
\newcommand{\bfl}{\begin{flushleft}}
\newcommand{\efl}{\end{flushleft}}
\newcommand{\edm}{\end{displaymath}}
\newcommand{\bdm}{\begin{displaymath}}
\def\bea{\begin{eqnarray}}
\def\eea{\end{eqnarray}}

\newcommand{\beqs}{\begin{subequations}}
\newcommand{\eeqs}{\end{subequations}}
\newcommand{\eeq}{\end{equation}}
\newcommand{\beq}{\begin{equation}}

\newcommand{\Eref}[1]{Eq.~(\ref{#1})}
\newcommand{\Sref}[1]{Sec.~\ref{#1}}
\newcommand{\Fref}[1]{Fig.~\ref{#1}}
\newcommand{\Tref}[1]{Table~\ref{#1}}
\newcommand{\cref}[1]{Ref.~\cite{#1}}

\newcommand\eqs[2]{Eqs.~(\ref{#1}) and (\ref{#2})}

\newcommand{\Gr}{\ensuremath{\widetilde{G}}}

\newcommand{\ftn}{\footnotesize}

\newcommand{\ssz}{\scriptsize}
\newcommand{\TeV}{{\mbox{\rm TeV}}}

\newcommand{\GeV}{{\mbox{\rm GeV}}}

\newcommand{\eV}{{\mbox{\rm eV}}}
\newcommand{\EeV}{{\mbox{\rm EeV}}}
\newcommand{\PeV}{{\mbox{\rm PeV}}}

\newcommand{\astroph}[1]{{\tt astro-ph/#1}}
\newcommand{\arxiv}[1]{{\tt arXiv:#1}}

\newcommand{\etal}{{\it et al.\/}}

\newcommand{\Vhi}{\ensuremath{V_{\rm HI}}}

\def\to{\rightarrow}

\def\lf{\left(}
\def\rg{\right)}

\def\llgm{\left\lgroup}
\def\rrgm{\right\rgroup}

\newcommand{\mP}{\ensuremath{m_{\rm P}}}
\newcommand{\Mgut}{\ensuremath{M_{\rm GUT}}}

\newcommand{\ld}{\ensuremath{\lambda}}

\newcommand{\kp}{\ensuremath{\kappa}}

\newcommand\vev[1]{\langle {#1} \rangle}
\newcommand{\Yb}{\ensuremath{Y_{B}}}
\newcommand{\Yg}{\ensuremath{Y_{\Gr}}}

\newcommand{\Khi}{\ensuremath{K}}

\newcommand{\lm}{\ensuremath{\lambda_\mu}}

\newcommand{\nsu}{\ensuremath{{N_{0}}}}
\newcommand{\ns}{\ensuremath{n_{\rm s}}}
\newcommand{\as}{\ensuremath{\alpha_{\rm s}}}
\newcommand{\As}{\ensuremath{A_{\rm s}}}
\newcommand{\Ns}{\ensuremath{N_{\star}}}

\newcommand{\wrh}{\ensuremath{w_{\rm rh}}}

\newcommand{\fp}{\ensuremath{f_{K}}}

\newcommand{\ks}{\ensuremath{k_\star}}

\newcommand{\hd}{{\ensuremath{H_d}}}
\newcommand{\hu}{{\ensuremath{H_u}}}

\newcommand{\Ve}{\ensuremath{{V}}}

\newcommand{\msn}{\ensuremath{\what m_{\rm \dph}}}

\newcommand{\Trh}{\ensuremath{T_{\rm rh}}}
\newcommand{\sg}{\ensuremath{\phi}}

\newcommand{\sgx}{\ensuremath{\phi_\star}}

\newcommand{\sgf}{\ensuremath{\phi_{\rm f}}}

\newcommand{\se}{\ensuremath{\widehat\phi}}
\newcommand{\see}{\ensuremath{\widehat \phi}}

\def\ve{\varepsilon}

\def\bbet{{\bar\beta}}
\def\al{{\alpha}}

\def\n{\bar{n}}

\def\Ka{K\"{a}hler potential}

\def\Km{K\"{a}hler manifold}
\def\Kaa{K\"{a}hler~}

\def\thb{{\bar\theta}}
\def\thn{{\theta_{\Phi}}}

\newcommand{\phc}{\ensuremath{\Phi}}
\newcommand{\phcb}{\ensuremath{\bar\Phi}}

\def\thb{{\bar\theta}}
\def\thn{{\theta_{\Phi}}}

\newcommand{\tr}{{\mbox{\sf\ssz T}}}

\newcommand{\diag}{\mbox{\sf diag}}

\newcommand{\dphi}{\ensuremath{\what{\delta\phi}}}

\newcommand{\dph}{\ensuremath{\delta\phi}}
\newcommand{\what}{\ensuremath{\widehat}}

\newcommand{\plk}{{\slshape Planck}}

\newcommand{\aS}{\ensuremath{{\rm a}_S}}
\newcommand{\Ald}{\ensuremath{A_\lambda}}
\newcommand{\am}{\ensuremath{{\rm a}_{\mu}}}

\newcommand{\ha}{{THI}}
\newcommand{\hb}{{THI}}

\newcommand{\tmd}{{TMI}}
\newcommand{\emd}{{EMI}}

\newcommand{\mgut}{\ensuremath{M_{\rm GUT}}}

\newcommand{\Dex}{\ensuremath{\Delta_{\rm \star}}}
\newcommand\vevi[1]{\langle {#1} \rangle_{\rm I}}

\newcommand\mtta[4]{\mbox{
$\llgm\bem #1 &#2 \cr #3& #4\eem\rrgm$}}

\newcommand\mtt[4]{\mbox{
$\llgm\bem #1 &#2 \cr #3& #4\eem\rrgm$}}

\newcommand{\Gsn}{\ensuremath{\what{\Gamma}_{\rm \dph}}}
\newcommand{\GNsn}{\ensuremath{\what{\Gamma}_{\dph\to N_i^c}}}
\newcommand{\Ghsn}{\ensuremath{\what{\Gamma}_{\dph\to H}}}

\newcommand{\rhni}{\ensuremath{N^c_i}}
\newcommand{\mrh[1]}{\ensuremath{M_{#1N^c}}}
\newcommand{\mD[1]}{\ensuremath{m_{#1\rm D}}}
\newcommand{\mn[1]}{\ensuremath{m_{#1\rm \nu}}}

\newcommand{\mgr}{\ensuremath{m_{3/2}}}
\newcommand{\mg}{{\ensuremath{M_{1/2}}}}

\begin{document}
\mainmatter              
\title{T-Model Higgs Inflation in Supergravity}
\titlerunning{T-Model HI in SUGRA}  
%
\author{Constantinos Pallis}
\authorrunning{C. Pallis} 
%
%
\institute{Laboratory of Physics, Faculty of Engineering,\\
Aristotle University of Thessaloniki,  GR-541 24 Thessaloniki,
GREECE \email{kpallis@gen.auth.gr}}

\maketitle              

\begin{abstract}

We focus on a simple, natural and predictive T model of inflation
in Supergravity employing as inflaton the Higgs field which leads
to the spontaneous breaking of a $U(1)_{B-L}$ symmetry at the SUSY
GUT scale. We use a renormalizable superpotential, fixed by a
$U(1)$ $R$ symmetry, and a \Ka\ which parameterizes the \Km\
$SU(2,1)/(SU(2)\times U(1))\times(SU(2)/U(1))$ with scalar
curvature ${\cal R}_{K}=-6/N+2/\nsu$ where $0<\nsu<6$. The
spectral index $\ns$ turns out to be close to its present central
observational value and the tensor-to-scalar ratio $r$ increases
with $N\lesssim36$. The model can be nicely linked to MSSM
offering an explanation of the magnitude of the $\mu$ parameter
consistently with phenomenological data. It also allows for
baryogenesis via non-thermal leptogenesis with gravitino as light
as $1~\TeV$.

\end{abstract}
\section{From Minimal to T-model Higgs
Inflation}\label{intro}

The identification of the inflaton -- i.e., the scalar field
driving inflation -- remains an important open issue in the
are(n)a of inflationary model building \cite{review}. We aspire to
identify $\sg$ with the radial component of a Higgs field,
$\phc=\phi e^{i\theta}$, within a \emph{Grand Unified Theory}
({\sf GUT}). We adopt the term \emph{Higgs inflation} ({\sf HI})
for the relevant models \cite{old} and we clarify that it is not
limited to those which employ as higgsflaton \cite{kaloper} the
\emph{Standard Model} ({\sf SM}) Higgs field \cite{sm}. We find it
technically convenient to exemplify HI in our talk taking as
reference theory an ``elementary" GUT based on the gauge group
$G_{B-L}= G_{\rm SM}\times U(1)_{B-L}$ -- where ${G_{\rm SM}}$ is
the gauge group of the SM and $B$, $L$ denote baryon and lepton
number respectively. Despite its simplicity, $G_{B-L}$ is strongly
motivated by neutrino physics and leptogenesis mechanisms -- see
e.g. \cref{univ,ighi}. Therefore, we choose as inflationary
potential the one employed for the realization of the Higgs
mechanism,
\beq V_{\rm
HI}(\sg)=\ld^2(\sg^2-M^2)^{2}/16\simeq\ld^2\sg^4/16\>\>\>\mbox{for}\>\>\>M\ll\mP=1.
\label{vhi}\eeq
If $\sg$ is canonically normalized, the theoretically derived
values $\ns\simeq0.947$ and $r\simeq0.28$ are not compatible with
the observational ones, since the combined {\sc Bicep2}/{\slshape
Keck Array} and \plk\ results require \cite{plin,gws}
\beq \ns=0.965\pm0.009~~~\mbox{and}~~~ r\lesssim0.032~~~\mbox{at
95\% c.l.}\label{data}\eeq

On the contrary, observationally friendly are models called
$\alpha$-attractors which employ chaotic potentials and so can be
activated with $\Vhi$ in \Eref{vhi}. These are based on the
specific relation established between the initial, $\sg$, and the
canonically normalized inflaton $\se$ and can be classified into
\emph{E-Model Inflation} ({\sf\small EMI}) \cite{alinde,epole} (or
$\alpha$-Starobinsky model \cite{eno7}) and \emph{T-Model
Inflation} ({\sf\small \tmd}) \cite{tmodel}. I.e.
\beq \phi=\begin{cases}1-{\rm Exp}\lf{-\sqrt{2/N}\se}\rg& \mbox{for \emd,}\\
\tanh{\lf\se/\sqrt{2N}\rg}& \mbox{for \tmd,}
\end{cases}~~\mbox{with}~~N>0.\eeq
Such relations between $\sg$ and $\se$ can be achieved in the
presence of a pole in the  inflaton kinetic term. TMI is ``taylor
made'' for HI, since it arises from a kinetic pole of order two
which includes the GUT-invariant quantity
$|\phc|^2:=\phc^\dagger\phc$. In particular, the lagrangian ${\cal
L}$ of $\sg=\sg(t)$ reads
\beq {\cal  L} = \sqrt{-\mathfrak{g}} \left({N}\dot\sg^2/{f_K^2}
-\Vhi(\sg)\right)~~\mbox{with}~~\dot{}=d/dt,~~f_K=1-\sg^2~~\mbox{and}~~N>0.\eeq
Here $t$ is the cosmic time, $\mathfrak{g}$ is the determinant of
the space-time metric $g_{\mu\nu}$ with signature $(+,-,-,-)$. If
we extract $\se$,  we obtain
\beq
\frac{d\se}{d\phi}=J=\frac{\sqrt{2N}}{f_K}~~\Rightarrow\>\>\sg=\tanh\frac{\se}{\sqrt{2N}}\>\>\>\mbox{and}\>\>\>
\Vhi(\se)\simeq
\frac{\ld^2}{16}\tanh^4\frac{\se}{\sqrt{2N}},\label{Jt}\eeq
by virtue of \Eref{vhi}. $\Vhi$ expressed as a function of $\se$
develops a plateau for $\se>1$ \cite{tmodel,sor} which renders it
convenient for the realization of a observationally viable
\emph{T-Model HI} ({\sf\small \hb}). The natural framework of a
GUT is \emph{Supersymmetry} ({\sf\small SUSY}) -- and its topical
extension, \emph{Supergravity} ({\sf\small SUGRA}) -- where the
gauge hierarchy problem is set under control. We below, in
\Sref{fhim}, we describe a SUGRA embedding -- from those
introduced in \cref{sor} -- of the model above and then, in
\Sref{secpost}, we propose a possible post-inflationary completion
which assures generation of the $\mu$ term of \emph{Minimal SUSY
SM} ({\sf\small MSSM}) and the \emph{baryon number of the
universe} ({\sf\small BAU}) via \emph{non-thermal leptogenesis}
({\sf\small nTL}).

\section{SUGRA Embedding}\label{fhim}

The part of the SUGRA lagrangian including the (complex) scalar
fields $z^\al$ can be written as
\beq\label{Saction1} {\cal  L} = \sqrt{-\mathfrak{g}} \lf
K_{\al\bbet} \partial_\mu z^\al \partial^\mu z^{*\bbet}-V_{\rm
F}+\cdots\rg, \eeq
where the kinetic mixing is controlled by the K\"ahler potential,
$K$, and the relevant metric defined as
\beq \label{kddef} K_{\al\bbet}={\Khi_{,z^\al
z^{*\bbet}}}>0\>\>\>\mbox{with}\>\>\>K^{\bbet\al}K_{\al\bar
\gamma}=\delta^\bbet_{\bar \gamma}.\eeq
We focus on the F-term part of the SUGRA scalar potential, $V_{\rm
F}$, which is given in terms of $K$, and the superpotential, $W$,
by
\beq \Ve_{\rm F}=e^{\Khi}\left(K^{\al\bbet}D_\al W D^*_\bbet
W^*-3{\vert W\vert^2}\right)~~\mbox{with}~~D_\al W=W_{,z^\al}
+K_{,z^\al}W \label{Vsugra} \eeq
the K\"ahler covariant derivative \emph{with respect to}
({\sf\small w.r.t}) the field $z^\al$ -- the symbol $,Z$ as
subscript denotes derivation w.r.t $Z$. The non-SUSY model
introduced in \Sref{intro} can be reproduced if we adopt
\cite{sor}
\beqs\bea W&=&\ld S\lf \bar\Phi\Phi-M^2/2\rg /2,\label{whi}\\
K&=&-N\ln\frac{1-|\phcb|^2-|\phc|^2}{\sqrt{(1-2\phcb\phc)(1-2\phcb^*\phc^*)}}+\nsu\ln\lf1+\frac{|S|^2}{\nsu}\rg,
\label{khi}\eea\eeqs
which are consistent with the $U(1)_{B-L}$ and an $R$ symmetry
acting on the various superfields as follows
\beq (B-L)(S, \phcb, \Phi)=(0,-2,2)~~~\mbox{and}~~~R(S, \phcb,
\Phi)=(1,0,0).\label{qass}\eeq
$W$ leads to a $B-L$ phase transition since the vacuum of the SUSY
limit of $\Ve_{\rm F}$ lies at the direction
\beq
\label{vevs}\vev{S}\simeq0~~\mbox{and}~~|\vev{\Phi}|=|\vev{\bar\Phi}|\simeq
M/\sqrt{2}~~\mbox{for}~~M\ll1,\eeq
%
i.e., $U(1)_{B-L}$ is spontaneously broken via the vacuum
expectation values of $\Phi$ and $\bar\Phi$. On the other hand,
$K$ with $0<\nsu<6$ assures a stabilization of the $S=0$ direction
\cite{rube} during THI without invoking higher order terms
\cite{su11}. Moreover, $K$ has no contribution to the exponent in
$V_{\rm F}$ of \Eref{Vsugra} -- cf. \cref{tkref} -- and $K$
parameterizes the \Km\ \cite{sor} \beq (SU(2,1)/(SU(2)\times
U(1))_{\phcb\phc}\times(SU(2)/U(1))_S\eeq
with constant scalar curvature ${\cal R}_{K}=-6/N+2/\nsu$ -- the
indices above indicate the moduli which parameterizes the
respective \Kaa submanifolds.

\subsection{Inflationary Potential and Observables}\label{fhi11}

Adopting for the relevant fields the parameterizations  \beq
\Phi=\phi e^{i\theta}\cos\theta_\Phi,\>\>\> \bar\Phi=\phi
e^{i\thb}\sin\theta_\Phi\>\>\>
\mbox{with}\>\>\>0\leq\thn\leq{\pi}/{2}~~~\mbox{and}~~~S= \frac{s
+i\bar s}{\sqrt{2}},\eeq
we can easily verify that a D-flat direction is \beq
\vevi{\theta}=\vevi{\thb}=0,\>\vevi{\thn}={\pi/4}\>\>\>\mbox{and}\>\>\>\vevi{S}=0,\label{inftr}\eeq
which can be qualified as inflationary path. Substituting it and
\eqs{khi}{whi} into \Eref{Vsugra}, $\Vhi$ in \Eref{vhi} can be
reproduced from the just one surviving term, $\vevi{|W_{,S}|^2}$.
To achieve \tmd, though, we have to establish the correct
non-minimal kinetic mixing shown in \Eref{Jt}. To this end, we
compute $K_{\al\bbet}$ along the path in \Eref{inftr} which takes
the form
\beq \vevi{K_{\al\bbet}}=\diag\lf
\vevi{M_{\phcb\phc}},\vevi{K_{SS^*}}\rg~~\mbox{with}~~
\vevi{M_{\phcb\phc}}=\frac{\kp\sg^2}{2}\mtta{2/\sg^2-1}{1}{1}{2/\sg^2-1},\eeq
$\kp={N}/{\fp^{2}}$ and $K_{SS^*}=1$. Then we diagonalize
$\vevi{M_{\phcb\phc}}$ via a similarity transformation as follows:
\beq U_{\phcb\phc} \vevi{M_{\phcb\phc}} U_{\phcb\phc}^\tr
=\diag\lf \kp_+,\kp_-\rg\>\>\>\mbox{where}\>\>\>U_{\phcb\phc}=
\frac{1}{\sqrt{2}}\mtt{1}{1}{-1}{1}, \eeq
$\kp_+=\kp$ and $\kp_-=\kp\fp$. Canonically normalizing the
various fields, we obtain the desired form of $J$ in \Eref{Jt}. We
can also verify that the direction of \Eref{inftr} is stable w.r.t
the fluctuations of the non-inflaton fields and determine $M$
demanding that the GUT scale $\Mgut\simeq2/2.433\times10^{-2}$ is
identified with the mass of $B-L$ gauge boson at the vacuum of
\Eref{vevs}, i.e. \cite{sor},
\beq \label{Mg} \vev{M_{BL}}={\sqrt{2N}gM/
\vev{f_K}}=\mgut\>\>\Rightarrow\>\>M\simeq{\mgut}/{g\sqrt{2N}}
\>\>\mbox{with}\>\>\>g\simeq0.7.\eeq

\begin{figure}[t]
\hspace*{-0.2cm}\begin{minipage}{95mm}
\includegraphics[height=8.5cm,angle=-90]{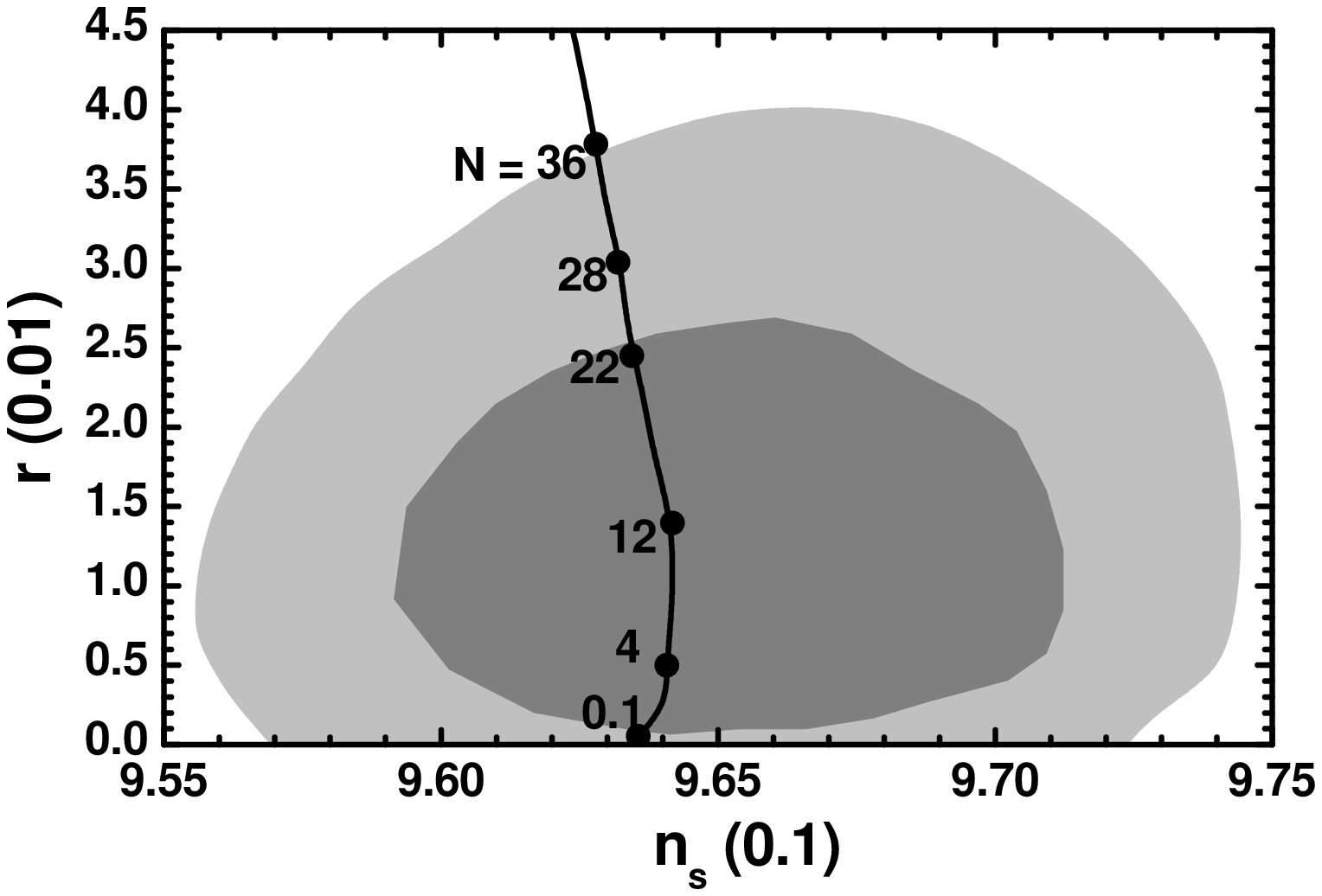}
\end{minipage}
\hfill
\begin{minipage}{25mm}
\renewcommand{\arraystretch}{1.1}
\begin{center}
{\small \begin{tabular}{cc}\hline $N$&$12$\\\hline
$\sgx/0.1$&$9.75$\\
$\Dex (\%)$&$2.5$\\
$\sgf/0.1$&$3.9$\\\hline
$\wrh$&$0.266$\\
$\Ns$&$56.4$\\\hline
$\ld/10^{-5}$&$8.6$\\\hline
$\ns/0.1$&$9.64$\\
$-\as/10^{-4}$&$6.4$\\
$r/10^{-2}$&$1.4$\\\hline
\end{tabular}}\renewcommand{\arraystretch}{1.}
\end{center}
\end{minipage}
\caption{\sl Curve allowed by \eqs{Mg}{Prob} in the $\ns-r$ plane
for various $N$'s indicated along it. The marginalized joint
$68\%$ [$95\%$] c.l. regions \cite{plin, gws} from PR4,
{\sffamily\small BK18}, BAO and lensing data-sets are depicted by
the dark [light] shaded contours.  The relevant field values,
parameters and observables corresponding to $N=12$ are listed in
the Table.} \label{fig2}\end{figure}

The allowed parameter space of our model can be delineated
enforcing the basic inflationary prerequisites \cite{plcp}, i.e.,
\beq \Ns\simeq55~~~\mbox{and}~~~\As\simeq4.588\cdot10^{-5},
\label{Prob} \eeq
where $\Ns$ is the number that the pivot scale $\ks=0.05/{\rm
Mpc}$ suffers during THI and $\As$ is the amplitude of the
curvature perturbation generated by the inflaton when $\ks$
crosses outside the inflationary horizon for $\sg=\sgx$. Enforcing
\Eref{Prob} we constrain $\sgx$ and $\ld$ and obtain the allowed
curve in the $\ns-r$ plane -- see \Fref{fig2} -- by varying $N$;
note that we use $\nsu=1$ throughout. A comparison with the
observational data \cite{plin,gws} is also displayed there
together with some typical values of the parameters for $N=12$. We
observe that $\ns$ and $r$ increase with $N$. More specifically,
we obtain
\beq \label{resi} 0.963\lesssim\ns\lesssim0.964,\>\>\>0.1\lesssim
N\lesssim 36\>\>\>\mbox{and}\>\>\> 0.0005\lesssim
{r}\lesssim0.039,\eeq
where $\wrh\simeq0.3$ is the barotropic index during the
inflaton-decaying domination and $\Ns\simeq56$. The proximity of
$\sgx$ to $1$ signals a mild tuning in the initial conditions
since $\Dex=1-\sgx\simeq(0.2-7)\%$ increasing with $N$.

\section{Post-Inflationary Completion}\label{secpost}

THI can be embedded in a $B-L$ extension of MSSM promoting to
gauge the pre-existing global $U(1)_{B-L}$. The terms of the total
super- and \Kaa\ potential which control the coexistence of the
inflationary and the MSSM sectors are
\beqs\bea \label{dW}\Delta W&= &\ld_{\mu} S\hu\hd\ +\ \ld_{ij\nu}
\bar\Phi N^{c}_iN^{c}_j,\\
\Delta
K&=&\mbox{$\sum_{a=1}^{5}$}\nsu\ln\lf1+|X_\al|^2/\nsu\rg~~\mbox{with}~~X_\al=N^{c}_i,\hu,\hd\label{dK}\eea\eeqs
where $\hu$ and $\hd$ are the electroweak Higgs superfields \&
$N^{c}_i$ the $i$th generation right-handed neutrinos with
$i=1,..,3$. These fields are supposed to be stabilized at zero
during THI, i.e. $\vevi{X^\al}=0$, and their charge assignments
are
\beq (B-L)(N^{c}_i, \hu, \hd)=(1,0,0)~~~\mbox{and}~~~R(N^{c}_i,
\hu, \hd)=(1,0,0).\eeq
The phenomenological consequences of the terms in \eqs{dW}{dK}
include the generation of the $\mu$ term and nTL explored in
Sec.~\ref{secmu} and \ref{seclepto} below. Hereafter we restore
units, i.e., we take $\mP=2.433\cdot10^{18}~\GeV$.

\subsection{$\mu$ Term and MSSM Phenomenology} \label{secmu}

The contributions from the soft SUSY-breaking terms, although
negligible during THI, may shift slightly $\vev{S}$ from zero in
\Eref{vevs}. Indeed, the relevant potential terms are
\beq V_{\rm soft}= \lf\ld A_\ld S \phcb\phc- {\rm a}_{S}S\ld M^2/4
+ {\rm h. c.}\rg+ m_{\gamma}^2\left|X^\gamma\right|^2,
\label{Vsoft} \eeq
where $m_{\gamma}\ll M, A_\ld$ and $\aS$ are soft SUSY-breaking
mass parameters and we assume $\hu=\hd=0$.  Rotating $S$ in the
real axis by an appropriate $R$-transformation, choosing
conveniently the phases of $\Ald$ and $\aS$ so as the total low
energy potential $V_{\rm tot}=V_{\rm SUSY}+V_{\rm soft}$ to be
minimized and substituting in $V_{\rm soft}$ the $\phc$ and
$\phcb$ values from \Eref{vevs} we get
\beqs\beq \vev{V_{\rm tot}(S)}= \ld^2\frac{M^2S^2}{4N} -\ld
M^2S\am \mgr ~~\mbox{with}~~~\am=\frac{|A_\ld| + |{\rm
a}_{S}|}{2\mgr}, \label{Vol} \eeq
where $\mgr$ is the gravitino ($\Gr$) mass and $\am>0$ is a
parameter of order unity which parameterizes our ignorance for the
dependence of $|A_\ld|$ and $|{\rm a}_{S}|$ on $\mgr$.  The
extermination condition for $\vev{V_{\rm tot}(S)}$ w.r.t $S$ leads
to a non vanishing $\vev{S}$ as follows
\beq \label{vevS}{d}\vev{V_{\rm tot}(S)}/{d S}
=0~~\Rightarrow~~\vev{S}=
{2N}\am\mgr/{\ld}\simeq{\sqrt{N}\Ns\am\mgr}/{2\pi\sqrt{6\As}},\eeq\eeqs
where we employed the $\ld-\As$ condition implied by the rightmost
expression in \Eref{Prob} -- see \cref{sor}. The generated $\mu$
parameter from the first term in \Eref{dW} is $\mu =\lm \vev{S}
\sim\mgr$ for $\lm\sim10^{-6}$. The achieved size of $\lm$ is
welcome, since stability of the $\hu-\hd$ system during \ha\
dictates -- cf. \cref{univ, ighi}:
\beq \lm<\ld(1+\nsu)\sgf^2/4\nsu\mP^2\sim10^{-5}\>\>\mbox{for}\>\>
\nsu=1.\label{lmb} \eeq
Thanks to the presence of $\sqrt{\As}\sim10^{-5}$ in the
denominator of the expression in \Eref{vevS}, almost any
$\mu/\mgr<1$ value is accessible for $\lm<10^{-5}$ without causing
any ugly hierarchy between $\mgr$ and $\mu$.

Adopting the \emph{Constrained MSSM} ({\sf\small CMSSM})
\cite{mssm}, as a low-energy framework, we can check its
compatibility with the proposed inflationary model doing some
simple estimates. Namely, CMSSM is relied on the following $4$ and
$1/2$ free parameters
\begin{equation}
{\rm
sign}\mu,~~\tan\beta=\vev{\hu}/\vev{\hd},~~\mg,~~m_0~~\mbox{and}~~A_0,
\label{para}
\end{equation}
where the three last mass parameters denote the common gaugino
mass, scalar mass and trilinear coupling constant, respectively,
defined (normally) at $\Mgut$. Imposing a number of
cosmo-phenomenological constraints the best-fit values of $|A_0|$,
$m_0$ and $|\mu|$ can be determined as in \cref{mssm}. Some of
their results are listed in the first four lines of \Tref{tab}. We
see that there are four allowed regions characterized by the
specific mechanism for suppressing the relic density of the
lightest neutralino ($\tilde\chi$) -- which turns out to be the
lightest SUSY particle -- to an acceptable level of \cref{plcp},
i.e., $\Omega_{\tilde\chi}h^2\lesssim0.12$ -- $\tilde\tau_1,
\tilde t_1$ and $\tilde \chi^\pm_1$ stand for the lightest stau,
stop and chargino eigenstate whereas $A$ and $H$ are the CP-odd
and the heavier CP-even Higgs bosons of MSSM respectively. The
proposed regions pass all the currently available LHC bounds
\cite{lhc} on the masses of the various sparticles and assure for
the mass of the lighter CP-even Higgs boson $m_h\simeq125~\GeV$.

\renewcommand{\arraystretch}{1.3}
\begin{table}[t!]
\caption{\sl  $\lm$ values compatible with the CMSSM best-fit
points} \bec{\small
\begin{tabular}{lc|c|c|c||c|c|c}\hline \multicolumn{2}{c|}{\ftn\sc
CMSSM
Region}&$|A_0|~(\TeV)$&$m_0~(\TeV)$&$|\mu|~(\TeV)$&$\am$&\multicolumn{2}{|c}{
$\lm$~({\boldmath $10^{-6}$})}\\\cline{7-8} &&&&&&$N=1$&$N=36$
\\\hline\hline
{\bfseries (I)}&$A/H$ Funnel &$9.9244$ &$9.136$&$1.409$&$1.086$ &{\boldmath $1.81$}&{\boldmath $0.3$}\\
{\bfseries (II)}&$\tilde\tau_1-\tilde\chi$ Coannihilation &$1.2271$ &$1.476$&$2.62$& $0.831$&{\boldmath ${\it 14.48}$}&{\boldmath $5$ }\\
{\bfseries (III)}&$\tilde t_1-\tilde\chi$ Coannihilation  &$9.965$ &$4.269$&$4.073$&$2.33$ &{\boldmath $5.2$}&{\boldmath $0.9$}\\
{\bfseries (IV)}&$\tilde \chi^\pm_1-\tilde\chi$ Coannihilation  &$9.2061$ &$9.000$&$0.983$&$1.023$ &{\boldmath $1.35$}&{\boldmath $0.2$}\\
\hline
\end{tabular}}\eec\label{tab}\end{table}

\renewcommand{\arraystretch}{1.}

Enforcing the conditions for the electroweak symmetry breaking, a
value for the parameter $|\mu|$ can be achieved in each of the
regions in \Tref{tab}. Taking this $|\mu|$ value as input we can
extract the $\lm$ values, if we first derive $\am$ setting, e.g.,
\beq
m_0=\mgr~~~\mbox{and}~~~|A_0|=|A_\ld|=|\aS|.\label{softass}\eeq
The outputs of our computation are listed  in the two rightmost
columns of \Tref{tab} in bold for $N=1$ and $N=36$. From these we
infer that the required $\lm$ values -- decreasing with $N$
increasing --, in all cases besides the one written in italics,
are comfortably compatible with \Eref{lmb}. As regards nTL, -- see
\Sref{seclepto} below -- the $\Gr$ constraint does not favor any
specific case since the unstable $\Gr$  with $\mgr\gtrsim1~\TeV$
turns out to be cosmologically safe within our framework.

\subsection{Non-Thermal Leptogenesis and Neutrino
Masses}\label{seclepto}

Soon after the end of THI, the (canonically normalized) inflaton
\beq\dphi=\vev{J}\dph\>\>\>\mbox{with}\>\>\> \dph=\phi-M
\>\>\>\mbox{and}\>\>\>\vev{J}=\sqrt{2N}\label{dphi} \eeq
acquires mass -- reducing with increasing $N$ in its domain of
\Eref{resi} -- given by
\beq \label{msn} %
\msn=\left\langle\Ve_{\rm HI,\se\se}\right\rangle^{1/2}=
\left\langle \Ve_{\rm HI,\sg\sg}/J^2\right\rangle^{1/2}=\frac{\ld
M}{2\sqrt{N}}\simeq(4.4-26)\cdot10~\EeV,\eeq
where $1~\EeV=10^9~\GeV$.  This value is lower than that
encountered in models of induced-gravity HI \cite{ighi} and close
to that obtained in several versions of non-minimal HI
\cite{univ}. Meantime, $\dphi$ settles into a phase of damped
oscillations abound the minimum in \Eref{vevs} reheating the
universe at a temperature \cite{ighi}
\beq\Trh=
\left({72/5\pi^2g_{*}}\right)^{1/4}\lf\Gsn\mP\rg^{1/2}\>\>\>\mbox{with}\>\>\>\Gsn=\GNsn+\Ghsn\,.\label{Trh}\eeq
Here $g_{*}=228.75$ counts the MSSM effective number of
relativistic degrees of freedom and we take into account the
following decay widths
\beqs\bea \GNsn&=&\frac{g_{iN^c}^2}{16\pi}\msn\lf1-\frac{4\mrh[
i]^2}{\msn^2}\rg^{3/2}\>\>\mbox{with}\>\>\>
g_{iN^c}=\frac{\ld_{iN^c}}{\vev{J}},\\
\Ghsn&=&\frac{2}{8\pi}g_{H}^2\msn\>\>\>\>\mbox{with}\>\>\>\>
g_{H}=\frac{\lm}{\sqrt{2}}. \eea\eeqs
For $\Trh<\mrh[i]$, the out-of-equilibrium decay of $\rhni$
generates a lepton-number-asymmetry yield which is partially
converted through sphaleron effects into a yield of the observed
BAU
\beq Y_B=-0.35\cdot{5\over2}{\Trh\over\msn}\mbox{$\sum_i$}
{\GNsn\over\Gsn}\ve_i\,.\label{Yb}\eeq
Here $\vev{\hu}\simeq174~\GeV$, for large $\tan\beta$, $\ve_i$ is
the lepton asymmetry generated per $\rhni$ decay and $m_{\rm D}$
is the Dirac mass matrix of neutrinos, $\nu_i$.  The validity of
\Eref{Yb} requires the decay of $\dphi$ into at least one pair of
$\rhni$ and the avoidance of any erasure of the produced lepton
asymmetry yield. These conditions can be fulfilled if we demand
\beq\label{kin}
\msn\geq2\mrh[1]\>\>\>\mbox{and}\>\>\>\mrh[1]\gtrsim 10\Trh.\eeq
Moreover, \Eref{Yb} has to reproduce the observational result
\cite{plcp}
\beq Y_B=\lf8.697\pm0.054\rg\cdot10^{-11}.\label{BAUwmap}\eeq
The required $\Trh$ in \Eref{Yb} must be compatible with
constraints on the  $\Gr$ abundance, $Y_{3/2}$, at the onset of
\emph{nucleosynthesis} ({\sf\small BBN}), which is estimated to be
\beq\label{Ygr} Y_{3/2}\simeq 1.9\cdot10^{-22}\ \Trh/\GeV,\eeq
where we take into account only thermal production of $\Gr$, and
assume that $\Gr$ is much heavier than the MSSM gauginos. On the
other hand, $\Yg$  is bounded from above in order to avoid
spoiling the success of the BBN. For the typical case where $\Gr$
decays with a tiny hadronic branching ratio, we have
\beq  \label{Ygb} \Yg\lesssim\left\{\bem
%
10^{-14}\hfill \cr
10^{-13}\hfill \cr \eem
\right.\>\>\>\>\mbox{for}\>\>\>\>\mgr\simeq\left\{\bem
0.69~\TeV\hfill \cr
10.6~\TeV\hfill \cr \eem
\right.\>\>\>\>\mbox{implying}\>\>\>\>\Trh\lesssim5.3\cdot\left\{\bem
%
10^{-2}~\EeV\,,\hfill \cr
10^{-1}~\EeV\,.\hfill \cr\eem
\right.\eeq
The bounds above can be somehow relaxed in the case of a stable
$\Gr$.

Confronting $Y_B$ and $\Yg$ -- see \eqs{Yb}{Ygr} -- with
observations we can constrain the parameters of the neutrino
sector as function of the inflationary parameters. Indeed, the
seesaw formula expresses the light-neutrino masses $m_{i\nu}$ in
terms of $\mD[i]$ -- involved in \Eref{Yb} -- and $\mrh[i]$,
generated by the second term in \Eref{dW}. We follow the bottom-up
approach detailed in \cref{ighi,univ}, according to which we find
the $\mrh[i]$'s by using as inputs the $\mD[i]$'s, a reference
mass $\mn[\rm r]$ of the $\nu_i$'s -- $\mn[1]$ for \emph{normal
ordered} ({\sf\small NO}) $\mn[i]$'s, or $\mn[3]$ for
\emph{inverted ordered} ({\sf\small  IO} $\mn[i]$'s --, the two
Majorana phases $\varphi_1$ and $\varphi_2$ of the PMNS matrix,
and the best-fit values for the neutrino oscillation parameters,
i.e., the differences of the masses-squared ($\Delta m^2_{21}$ and
$\Delta m^2_{31}$), the mixing angles ($\theta_{13}, \theta_{23}$
and $\theta_{12}$) and the Dirac phase $\delta$ of PMNS matrix.
Namely, we take into account the recently updated best-fit values
\cite{valle} according to which $\Delta m^2_{21}=7.5\cdot
10^{-5}\eV^2$ and $\sin^2\theta_{12}=0.318$ for both $\mn[i]$
orderings. Also, we take
\begin{itemize}

\item For NO $\mn[i]$'s, $\Delta m^2_{31}=2.55\cdot 10^{-3}\eV^2$,
$\sin^2\theta_{13}=0.022$, $\sin^2\theta_{23}=0.574$ and
$\delta=1.08\pi$.

\item For IO $\mn[i]$'s, $\Delta m^2_{31}=2.45\cdot 10^{-3}\eV^2$,
$\sin^2\theta_{13}=0.0225$, $\sin^2\theta_{23}=0.578$ and
$\delta=1.58\pi$.

\end{itemize}
Furthermore, the sum of $\mn[i]$'s is bounded from above at 95\%
c.l. by the data \cite{plcp, valle} as follows
\beq\mbox{$\sum_i$} \mn[i]\leq0.23~{\eV}~\mbox{for NO
$\mn[i]$'s}~~~\mbox{or}~~~\mbox{$\sum_i$}
\mn[i]\leq0.15~{\eV}~\mbox{for IO $\mn[i]$'s}.\label{sumnu}\eeq
%


\begin{figure}[!t]
\hspace*{-.05in}
\begin{minipage}{8in}
\epsfig{file=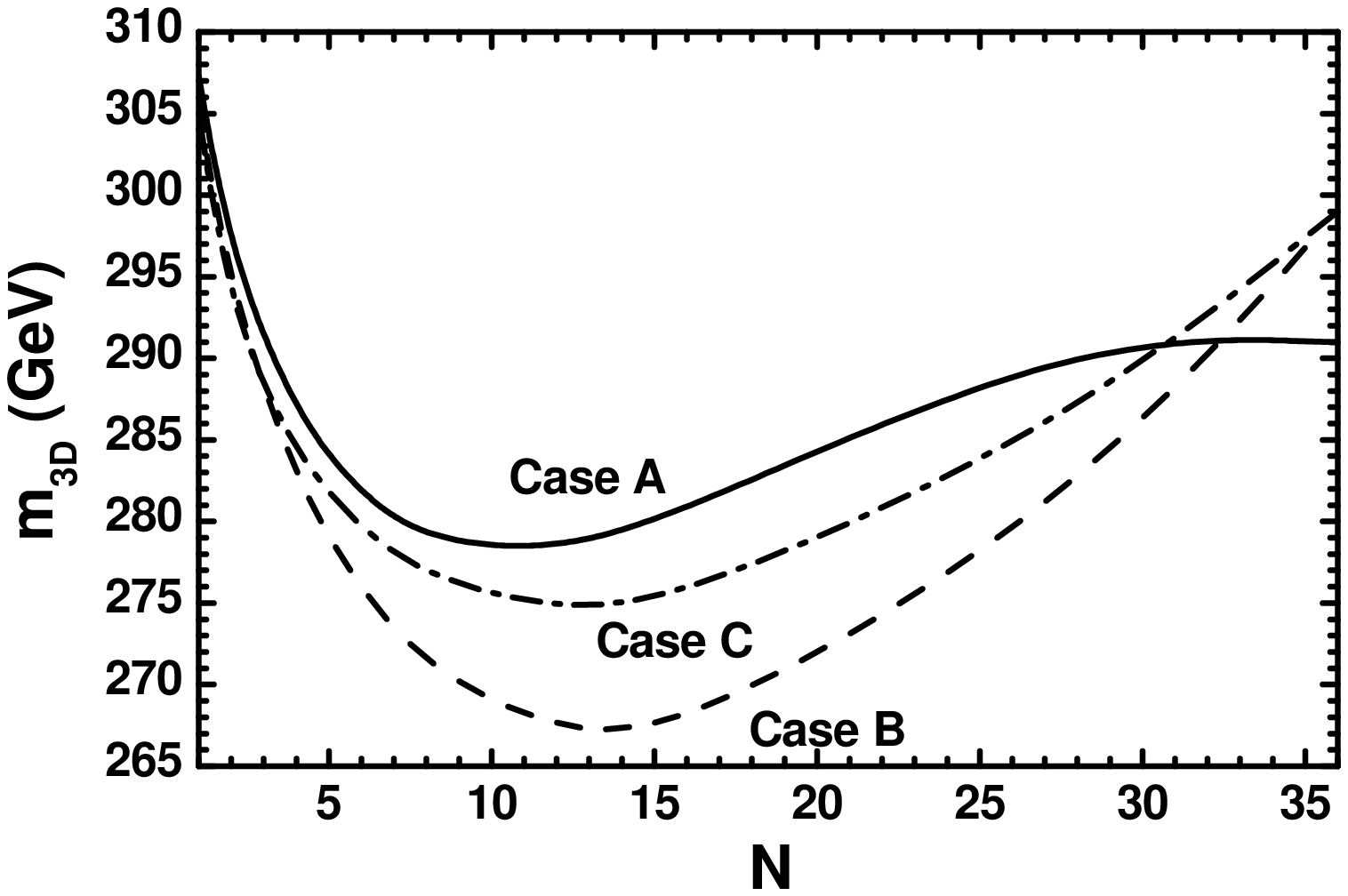,height=2.7in,angle=-90} \hspace*{-1.2cm}
\epsfig{file=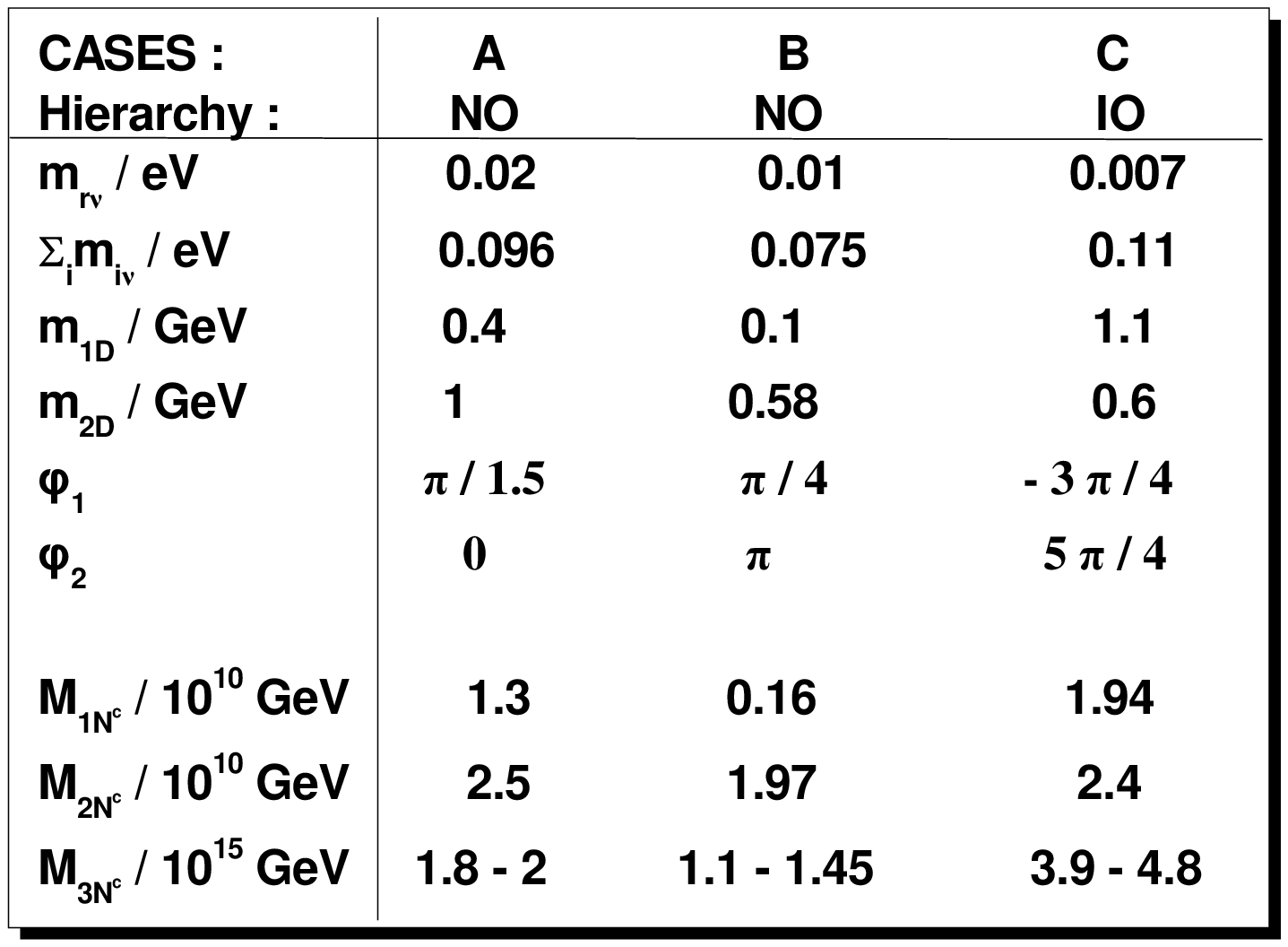,height=2.7in,angle=-90} \hfill
\end{minipage}
\hfill \caption{\sl\small  Contours, yielding the central $Y_B$ in
\Eref{BAUwmap} consistently with the inflationary requirements, in
the $N-m_{\rm 3D}$ plane. We take as inputs $\nsu=1$, $\mu=4~\TeV$
and the values of $m_{i\nu}$, $m_{\rm 1D}$, $m_{\rm 2D}$,
$\varphi_1$ and $\varphi_2$ shown in the Table which correspond to
cases A (solid line), B (dashed line) and C (dot-dashed
line).}\label{fmD}
\end{figure}

The outcome of our computation is presented in \Fref{fmD}, where
we depict the allowed values of $m_{\rm 3D}$ versus $N$ for
$\nsu=1$, $\mu=4~\TeV$ -- which corresponds to
$0.9\lesssim\lm/10^{-6}\lesssim5.2$ as $N$ ranges from $36$ to $1$
-- and the remaining parameters shown in the Table of \Fref{fmD}.
We use solid, dashed and dot-dashed line when the remaining inputs
-- i.e. $\mn[i]$, $\mD[3]$, $\varphi_1$, and $\varphi_2$ --
correspond to the cases A, B and C of the Table of \Fref{fmD}
respectively. We consider NO (cases A and B) and IO (case C)
$\mn[i]$'s. In all cases, the current limit in \Eref{sumnu} is
safely met. The gauge symmetry considered here does not predict
any particular Yukawa unification pattern and so, the $\mD[i]$'s
are free parameters.  Care is also taken so that the
perturbativity of $\ld_{iN^c}$ holds, i.e.,
$\ld_{iN^c}^2/4\pi\leq1$. The inflaton $\dphi$ decays mostly into
$N_1^c$'s. In all cases $\GNsn<\Ghsn$ and so the ratios
$\GNsn/\Gsn$ in \Eref{Yb} introduce a considerable reduction in
the derivation of $\Yb$. For the considered cases in \Fref{fmD},
successful nTL requires $\mrh[i]$ and $\mD[3]$ in the ranges
$(1-10^{6})~\EeV$ -- see Table of \Fref{fmD} -- and
$(265-308)~\GeV$ respectively. Meantime, $Y_{\Gr}$ and $\Trh$ in
\eqs{Trh}{Ygr} are confined in the ranges
\beq
1.6\lesssim\Trh/{10~\PeV}\lesssim23.3\>\>\>\mbox{and}\>\>\>3.2\lesssim
Y_{\Gr}/10^{-15}\lesssim44.3,\label{res}\eeq
with both quantities decreasing with $N$ increasing. Therefore,
the $\Gr$ constraint can be satisfied even for $\mgr \sim 1~\TeV$
as deduced comparing \eqs{Ygb}{res}. As advertised in
\Sref{secmu}, these values are in nice agreement with the ones
needed for the solution of the $\mu$ problem within CMSSM in all
cases of \Tref{tab}.

\section{Summary}

We focused on the simplest among the implementations introduced in
\cref{sor} of T-model inflation in SUGRA, using as inflaton a
Higgs field which breaks a gauge $U(1)_{B-L}$. We employed the
super- and \Kaa\ potentials $W$ and $K$ shown in \eqs{whi}{khi}
which lead to the inflationary potential in \Eref{vhi} and the
$\sg-\see$ relation in \Eref{Jt}. The model predicts
$\ns\sim0.963$ and $r$ increasing with the prefactor $N\lesssim36$
in \Eref{khi} without unnatural tuning. We also specified a
post-inflationary completion, based on the extra terms in
\eqs{dW}{dK}, which offers a nice solution to the $\mu$ problem of
MSSM and allows for baryogenesis via nTL with $\mrh[i]$ in the
range $(1-10^{6})~\EeV$.

\newpage

\paragraph*{\small \bf\scshape Acknowledgments} {\small I would like to thank
N. Mavromatos, Q. Shafi and F. Farakos for interesting discussions
and questions. This research work was supported by the Hellenic
Foundation for Research and Innovation (H.F.R.I.) under the
``First Call for H.F.R.I. Research Projects to support Faculty
members and Researchers and the procurement of high-cost research
equipment grant'' (Project Number: 2251).}

\def\ijmp#1#2#3{{\emph{Int. Jour. Mod. Phys.}}
{\bf #1},~#3~(#2)}
\def\plb#1#2#3{{\emph{Phys. Lett.  B }}{\bf #1},~#3~(#2)}
\def\zpc#1#2#3{{Z. Phys. C }{\bf #1},~#3~(#2)}
\def\prl#1#2#3{{\emph{Phys. Rev. Lett.} }
{\bf #1},~#3~(#2)}
\def\rmp#1#2#3{{Rev. Mod. Phys.}
{\bf #1},~#3~(#2)}
\def\prep#1#2#3{\emph{Phys. Rep. }{\bf #1},~#3~(#2)}
\def\prd#1#2#3{{\emph{Phys. Rev.  D }}{\bf #1},~#3~(#2)}
\def\npb#1#2#3{{\emph{Nucl. Phys.} }{\bf B#1},~#3~(#2)}
\def\npps#1#2#3{{Nucl. Phys. B (Proc. Sup.)}
{\bf #1},~#3~(#2)}
\def\mpl#1#2#3{{Mod. Phys. Lett.}
{\bf #1},~#3~(#2)}
\def\arnps#1#2#3{{Annu. Rev. Nucl. Part. Sci.}
{\bf #1},~#3~(#2)}
\def\sjnp#1#2#3{{Sov. J. Nucl. Phys.}
{\bf #1},~#3~(#2)}
\def\jetp#1#2#3{{JETP Lett. }{\bf #1},~#3~(#2)}
\def\app#1#2#3{{Acta Phys. Polon.}
{\bf #1},~#3~(#2)}
\def\rnc#1#2#3{{Riv. Nuovo Cim.}
{\bf #1},~#3~(#2)}
\def\ap#1#2#3{{Ann. Phys. }{\bf #1},~#3~(#2)}
\def\ptp#1#2#3{{Prog. Theor. Phys.}
{\bf #1},~#3~(#2)}
\def\apjl#1#2#3{{Astrophys. J. Lett.}
{\bf #1},~#3~(#2)}
\def\n#1#2#3{{Nature }{\bf #1},~#3~(#2)}
\def\apj#1#2#3{{Astrophys. J.}
{\bf #1},~#3~(#2)}
\def\anj#1#2#3{{Astron. J. }{\bf #1},~#3~(#2)}
\def\mnras#1#2#3{{MNRAS }{\bf #1},~#3~(#2)}
\def\grg#1#2#3{{Gen. Rel. Grav.}
{\bf #1},~#3~(#2)}
\def\s#1#2#3{{Science }{\bf #1},~#3~(#2)}
\def\baas#1#2#3{{Bull. Am. Astron. Soc.}
{\bf #1},~#3~(#2)}
\def\ibid#1#2#3{{\it ibid. }{\bf #1},~#3~(#2)}
\def\cpc#1#2#3{{Comput. Phys. Commun.}
{\bf #1},~#3~(#2)}
\def\astp#1#2#3{{Astropart. Phys.}
{\bf #1},~#3~(#2)}
\def\epjc#1#2#3{{Eur. Phys. J. C}
{\bf #1},~#3~(#2)}
\def\nima#1#2#3{{Nucl. Instrum. Meth. A}
{\bf #1},~#3~(#2)}
\def\jhep#1#2#3{{\emph{J. High Energy Phys.} }
{\bf #1},~#3~(#2)}
\def\jcap#1#2#3{{\emph{J. Cosmol. Astropart. Phys.} }
{\bf #1},~#3~(#2)}
\def\jcapn#1#2#3#4{{\sl J. Cosmol. Astropart. Phys. }{\bf #1}, no. #4, #3 (#2)}
\def\prdn#1#2#3#4{{\sl Phys. Rev. D }{\bf #1}, no. #4, #3 (#2)}
\def\prdn#1#2#3#4{{\sl Phys. Rev. D }{\bf #1}, no. #4, #3 (#2)}
\def\jcapn#1#2#3#4{{\sl J. Cosmol. Astropart.
Phys. }{\bf #1}, no. #4, #3 (#2)}
\def\epjcn#1#2#3#4{{\sl Eur. Phys. J. C }{\bf #1}, no. #4, #3 (#2)}

%
%

\end{document}